\documentclass[12pt]{iopart}
\usepackage{epsfig,bm}
\def\bbox#1{\hbox{\boldmath${#1}$}}
\begin{document}

\title{Explicit Solution of the Time Evolution of the Wigner
Function$^\dag$ }

{\footnotetext[1] {Invited talk presented at the Wigner Centennial
Conference, Pecs, Hungary, July 8-12, 2002, to be published in the
Journal of Optics B: Quantum and Classical Optics, June 2003.}  }

\author{Cheuk-Yin Wong
}

\address{Physics Division, Oak Ridge National Laboratory, Oak Ridge,
TN 37831-6373, USA
}

\begin{abstract}
Previously, an explicit solution for the time evolution of the Wigner
function was presented in terms of auxiliary phase space coordinates
which obey simple equations that are analogous with, but not identical
to, the classical equations of motion.  They can be solved easily and
their solutions can be utilized to construct the time evolution of the
Wigner function.  In this paper, the usefulness of this explicit
solution is demonstrated by solving a numerical example in which the
Wigner function has strong spatial and temporal variations as well as
regions with negative values.  It is found that the explicit solution
gives a correct description of the time evolution of the Wigner
function.  We examine next the pseudoparticle method which uses
classical trajectories to evolve the Wigner function.  We find that
the lowest-order pseudoparticle approximation reproduces the general
features of the time evolution, but there are deviations.  We show how
these deviations can be systematically reduced by including
higher-order correction terms in powers of $\hbar^2$.
\end{abstract}



\maketitle

\section{Introduction}
It is a great pleasure for me to participate in the Wigner Centennial
Conference at Pecs, Hungary in honor of the Centennial of Professor
Eugene Wigner's birthday.  I first met Professor Wigner in 1959 when I
was an undergraduate student at Princeton.  Professor Wigner taught us
and our fellow graduate school classmates Curtis Callen, Stephen Adler, and
Alfred Goldhaber a course in Advanced Quantum Mechanics in 1961.  We
were at the celebration party at the Graduate College when Professor
Wigner was awarded the Nobel Prize in Physics in 1963.  After I
obtained my Ph.\ D.\ degree at Princeton, I came to work at Oak Ridge
National Laboratory, for which Professor Wigner was the founding
Scientific Director.  I saw Professor Wigner from time to time at Oak
Ridge when he came to work on his civil defense project.  It is
therefore particularly meaningful to me to come and celebrate his
centennial in his native land to commemorate his many important
contributions.

As a part of the ``Junior Paper'' research project at Princeton
University, I went to see Professor Wigner in his office in Fine Hall
in 1959 and inquired about the early history of Quantum Mechanics.  He
told me that the great thing about Schr\"odinger was that his wave
equation was formulated in configuration space.  However, to
Professor Wigner, who was trained as a chemical engineer, dynamics
could also be described in phase space.  His formulation of
quantum mechanics in phase space in 1932 led to the well-known
Wigner description of quantum systems \cite{Wig32}.  The joint
function of coordinate and momentum $f(\bbox{r}\bbox{p})$ introduced
in 1932, now known as the Wigner function, is analogous with the
classical distribution function.  This analogy has provided new
insights and useful applications to many quantum systems.  It is
deservedly a subject of intense interest in this Wigner Centennial
Conference \cite{Wig03}.

The Wigner function is, however, not identical to the classical
distribution function.  The classical distribution function is always
non-negative and can be described in terms of a collection of cell
points with positive weights in phase space.  The evolution of the
classical distribution function can be followed by tracking these cell
points using classical equations of motion \cite{Won82}.

The Wigner function can assume negative as well as positive values in
some regions of phase space.  How does a Wigner function with regions
of negative values evolve as a function of time?  The concept of a
probability distribution function with positive weights is clearly not
applicable here.  The difficulty of propagating the Wigner function
with negative values has been a barrier to the study of the dynamics
of the Wigner function using the equation of motion of the Wigner
function directly.

Previously, I formulated a simple method to propagate the Wigner
function in time \cite{Won82}.  The ability to propagate the Wigner
function with negative values arises from the fact that one deals with
amplitudes in this method, and one does not need to use the concept of
probabilities.  The method contains three important components.
First, one employs an auxiliary variable $\bbox{s}$, which is allowed
to span the whole configuration space.  Second, using this auxiliary
variable $\bbox{s}$, the Wigner function can be represented in terms
of auxiliary phase space coordinates $\bbox{R}$ and $\bbox{P}$.  The
equations of motion for these auxiliary coordinates $\bbox{R}$ and
$\bbox{P}$ are quite simple.  They are analogous with, but not
identical to, the classical equations of motion, as $\partial\bbox{
P}/\partial t$ depends on the auxiliary coordinate $\bbox{s}$.  They
can be solved easily and their solutions can be utilized to construct
the Wigner function at the next time step.  The third component
consists of providing the correct amplitude function for the
evolution.  For each value of $\bbox{s}$ and the evolution of phase
space coordinates, one associates an amplitude factor $\exp\{i
\bbox{s}\cdot(\bbox{p}-\bbox{P})/\hbar \}$. After the amplitudes for
all possible values of $\bbox{s}$ have been obtained at the end of one
time step, one adds all these amplitudes together coherently (Huygen
Principle) to give the Wigner function $f(\bbox{r} \bbox{p})$ at the
next time step.  This explicit solution overcomes the difficulty of
propagating a Wigner function with negative values.

Recently, with the rapid advances in quantum information technology
and the reduction in the size of micro-electronic devices, there has
been renewed interest in experimental and theoretical studies of the
Wigner function \cite{Smi93}-\cite{Fed01}.  Experimental measurements
of the Wigner function have been carried out using many different
techniques \cite{Smi93}-\cite{Lou02}.  Regions of negative Wigner
function were observed in many experiments \cite{Lei96,Lvo01}.  The
time evolution of the Wigner function has been studied by many authors
\cite{Joh87}$-$\cite{Sch96}.  It is instructive to review the explicit
solution for the time evolution of the Wigner function obtained
previously and to demonstrate its usefulness by a numerical example.
We choose to examine an example in which the time evolution of the
Wigner function can be ready evaluated by using wave functions and
eigenvalues.  The Wigner function in the example should have
significant spatial and temporal variations as well as regions of
negative values to test whether the explicit solution obtained
previously is capable of treating these non-classical features.  The
success in providing a correct time evolution of the Wigner function
using this explicit solution will pave the way for its applications in
the future.

As an approximation of the explicit solution, a pseudoparticle
method was developed previously \cite{Won82} to obtain the time
evolution of the Wigner function by following the trajectories of the
phase space coordinates, using classical equations of motion .  It is
instructive to review this pseudoparticle method and assess its
accuracy by comparing with the correct results.  Furthermore, it is
useful to develop systematic ways to improve the accuracy of the
pseudoparticle method to assist its future applications.

\section {Explicit Solution of the Time Evolution of the Wigner
Function}

We shall first review the explicit solution obtained in 1982
\cite{Won82}. We consider a particle in the potential $V$
\begin{eqnarray}
i \hbar { \partial \over \partial t} \psi (\bbox{r},t)
=\left \{ - {\hbar ^2 \over 2m} \nabla^2 +{ V} (\bbox{r},t) \right \}
\psi (\bbox{r},t).  
\end{eqnarray}
The Wigner function is
\begin{eqnarray}
f(\bbox{r} \bbox{p}, t) = \int d\bbox{s} \, e^{i {\bbox p}\cdot {\bbox
s}/\hbar } \psi (\bbox{r}-{\bbox{s} \over 2},t)
    \psi^* (\bbox{r}+{ \bbox{s} \over 2},t).
\end{eqnarray} 
From the Schr\"odinger equation, we obtain the equation of motion for
the Wigner function \cite{Wig32}
\begin{eqnarray}
\label{dfdt}
{\partial f (\bbox{r} \bbox{p},t) \over dt} +{\bbox{p}\over m}\cdot
\nabla_r f(\bbox{r} \bbox{p},t) -{2 \over \hbar} \sin\left \{ {\hbar
\over 2} \nabla_r^V \cdot \nabla_p^f \right \} V(\bbox{r},t) f
(\bbox{r} \bbox{p},t) =0,
\end{eqnarray} 
where $\nabla_r^V$ acts on the potential ${ V}$ and $\nabla_p^f$ acts
on the Wigner function $f$.

Given the Wigner function $f (\bbox{r}_0 \bbox{p}_0,t_0)$ at time
$t_0$, we wish to obtain the Wigner function at the next time step at
$t=t_0+\delta t$ with a small $\delta t$.  To solve for the dynamics,
we divide the phase space into cells (pseudoparticles).  Consider one
such cell centered at $\{ \bbox{r}_0$ $\bbox{p}_0 \}$ at time $t_0$
with volume element $d{\bbox{r}_0} d {\bbox{p}_0}$ and amplitude $f
(\bbox{r}_0 \bbox{p}_0,t_0)$.  We retain the auxiliary coordinate
$\bbox{s}$ and follow the time dependence of each cell in the
Lagrangian sense.  For the phase space coordinates $\{\bbox{r}_0
\bbox{p}_0\}$ initially at time $t_0$, we label their phase space
coordinates at subsequent time $t$ by $\{\bbox{R}(\bbox{r}_0
\bbox{p}_0 \bbox{s},t)~ \bbox{P}(\bbox{r}_0 \bbox{p}_0 \bbox{s},t)\}$.
In classical dynamics, the evolution of the phase space coordinates
will not depend on $\bbox{s}$.  In quantum dynamics, they will depend
on $\bbox{s}$.  The momentum coordinate $\bbox{P}$ can jump to the
momentum coordinate $\bbox{p}$ at time $t$.  We associate an amplitude
of $\exp\{ i \bbox{s}\cdot[\bbox{p}-\bbox{P}(\bbox{r}_0 \bbox{p}_0
\bbox{s},t)]/\hbar\}$ for this momentum jump.  This amplitude factor
is chosen because for the classical case where $\bbox{P}$ is
independent of $\bbox{s}$ this amplitude leads to the correct delta
function propagator
$\delta(\bbox{p}-\bbox{P(\bbox{r}_0\bbox{p_0},t}))$ after we integrate
over $\bbox{s}$ [see Eq.\ (\ref{clasd})].  We therefore express the
Wigner function $ f (\bbox{r} \bbox{p},t)$ at the next time step as
\cite{Won82}
\begin{eqnarray}
\label{frp}
\fl
~~ f (\bbox{r} \bbox{p},t) 
= { \int} { d{\bbox{r}_0} d {\bbox{p}_0} {d\bbox{s}} \over (2 \pi
\hbar)^3}
\exp\{ i \bbox{s}\cdot[\bbox{p}-\bbox{P}
(\bbox{r}_0 \bbox{p}_0 \bbox{s},t)]/\hbar\}
\delta [ \bbox{r}-\bbox{R}(\bbox{r}_0 \bbox{p}_0 \bbox{s},t)]
f (\bbox{r}_0 \bbox{p}_0,t_0).
\end{eqnarray}
Our task is to find solutions for the phase space coordinates
$\bbox{R}(\bbox{r}_0 \bbox{p}_0 \bbox{s},t)$ and $\bbox{P}(\bbox{r}_0
\bbox{p}_0 \bbox{s},t)$ with initial conditions $\bbox{R}(\bbox{r}_0
\bbox{p}_0 \bbox{s},t_0)=\bbox{r}_0$ and $\bbox{P}(\bbox{r}_0
\bbox{p}_0 \bbox{s},t_0)=\bbox{p}_0$.  If we can find these
coordinates at the next time step, the above integral can then be
carried out to give the Wigner function at the new time $t$.

To obtain the equations of motion for $\bbox{R}(\bbox{r}_0 \bbox{p}_0
\bbox{s},t)$ and $\bbox{P}(\bbox{r}_0 \bbox{p}_0 \bbox{s},t)$, we
substitute Eq.\ (\ref{frp}) into Eq.\ (\ref{dfdt}).  For the first
term in Eq.\ (\ref{dfdt}), we get
\begin{eqnarray}
\label{1st}
\fl
~~{\partial f (\bbox{r} \bbox{p},t)  \over  \partial t}
= { \int} { d{\bbox{r}_0} d {\bbox{p}_0} {d\bbox{s}} \over (2 \pi
\hbar)^3}
f (\bbox{r}_0 \bbox{p}_0,t_0) \nonumber\\
\fl
~~~~~~~\times \Biggl [
{(-i \bbox{s}) \over \hbar} \cdot {\partial \bbox{P} \over  \partial t}
\exp\{ i \bbox{s}\cdot[\bbox{p}-\bbox{P}
(\bbox{r}_0 \bbox{p}_0 \bbox{s},t)]/\hbar\}
\delta [ \bbox{r}-\bbox{R}(\bbox{r}_0 \bbox{p}_0 \bbox{s},t)]
\nonumber\\
\!\!\!\!\!\!\!\!\!\!\!\!
+\exp\{ i \bbox{s}\cdot[\bbox{p}-\bbox{P}
(\bbox{r}_0 \bbox{p}_0 \bbox{s},t)]/\hbar\}
(\nabla_{\bbox{R}} 
\delta [ \bbox{r}-\bbox{R}(\bbox{r}_0 \bbox{p}_0 \bbox{s},t)])
\cdot \partial \bbox{R}(\bbox{r}_0 \bbox{p}_0 \bbox{s},t) /\partial t\Biggr ].
\end{eqnarray}
Noting that 
\begin{eqnarray}\nabla_{\bbox{R}} 
\delta [ \bbox{r}-\bbox{R}(\bbox{r}_0 \bbox{p}_0 \bbox{s},t)]
=
-\nabla_{\bbox{r}} 
\delta [ \bbox{r}-\bbox{R}(\bbox{r}_0 \bbox{p}_0 \bbox{s},t)],
\end{eqnarray}
we can rewrite the second term inside the square bracket as
\begin{eqnarray}
\fl
~~{ \int} { d{\bbox{r}_0} d {\bbox{p}_0} {d\bbox{s}} \over (2 \pi
\hbar)^3}
f (\bbox{r}_0 \bbox{p}_0,t_0) \nonumber\\
\fl
~~~~~~\times\exp\{ i \bbox{s}\cdot[\bbox{p}-\bbox{P}
(\bbox{r}_0 \bbox{p}_0 \bbox{s},t)]/\hbar\}
(-\nabla_{\bbox{r}} 
\delta [ \bbox{r}-\bbox{R}(\bbox{r}_0 \bbox{p}_0 \bbox{s},t)]
\cdot \partial \bbox{R}(\bbox{r}_0 \bbox{p}_0 \bbox{s},t) /\partial t 
 \nonumber \\
\fl
~~~~~~=
-{\partial \bbox{R}(\bbox{r}_0 \bbox{p}_0 \bbox{s},t) \over \partial t} 
\cdot 
\nabla_{\bbox{r}}
f (\bbox{r} \bbox{p},t),
\end{eqnarray}
where we have used Eq.\ (\ref{frp}) to obtain the right-hand side.
For the last term in Eq.\ (\ref{dfdt}), we substitute (\ref{frp}) into
this term and find
\begin{eqnarray}
\label{last}
\fl
~~{2 \over \hbar} \sin\left \{ {\hbar \over 2} \nabla_r^V \cdot
\nabla_p^f \right \} V(\bbox{r},t) f (\bbox{r} \bbox{p},t) 
={1 \over \hbar} { \int} { d{\bbox{r}_0} d {\bbox{p}_0} {d\bbox{s}}
\over (2 \pi \hbar)^3} f (\bbox{r}_0 \bbox{p}_0,t_0) \nonumber\\
\fl
~~~~~~
\times \exp\{ i \bbox{s}\cdot[\bbox{p}-\bbox{P}
(\bbox{r}_0 \bbox{p}_0 \bbox{s},t)]/\hbar\} 
\delta
[ \bbox{r}-\bbox{R}(\bbox{r}_0 \bbox{p}_0 \bbox{s},t)]) 
{[V(\bbox{r}-{\bbox{s} \over 2},t) 
-V(\bbox{r}+{\bbox{s} \over 2},t) ] \over i}.
\end{eqnarray}
Putting Eqs.\ (\ref{1st})-(\ref{last}) into Eq.\ (\ref{dfdt}), we
obtain
\begin{eqnarray}
\fl
~~\left [-{\partial \bbox{R}(\bbox{r}_0 \bbox{p}_0 \bbox{s},t) 
\over \partial t} 
+{\bbox{p}\over m} \right ]\cdot
\nabla_r f(\bbox{r} \bbox{p},t) 
\nonumber\\
+  { \int} { d{\bbox{r}_0} d {\bbox{p}_0} {d\bbox{s}} \over (2 \pi
\hbar)^3}
f (\bbox{r}_0 \bbox{p}_0,t_0)
\Biggl [
{(-i \bbox{s}) \over \hbar} 
\cdot {\partial \bbox{P} \over  \partial t}
- {[V(\bbox{r}-{\bbox{s} \over 2},t) 
-V(\bbox{r}+{\bbox{s} \over 2},t) ] \over i\hbar} \Biggr ]
\nonumber \\
\times \exp\{ i \bbox{s}\cdot[\bbox{p}-\bbox{P}
(\bbox{r}_0 \bbox{p}_0 \bbox{s},t)]/\hbar\}
\delta [ \bbox{r}-\bbox{R}(\bbox{r}_0 \bbox{p}_0 \bbox{s},t)]=0.
\end{eqnarray}
By comparing similar terms, the above equation leads to the equations
of motion for $\bbox{R}$ and $\bbox{P}$ \cite{Won82},
\begin{eqnarray}
\label{dR}
{\partial \bbox{R} \over \partial t} = {\bbox p \over m},
\end{eqnarray}
\begin{eqnarray}
\label{dP}
\bbox{s}\cdot {\partial \bbox{P} \over \partial t}
={ V} \left ( \bbox {R}- {\bbox{s} \over 2}, t \right ) 
-{ V} \left ( \bbox {R}+ {\bbox{s} \over 2}, t \right ) .
\end{eqnarray}
In the above equations, $\bbox{r}$ and $\bbox{R}$ are interchangeable,
because of the $\delta$-function $\delta(\bbox{r}-\bbox{R})$ in
Eq. (\ref{frp}).  It is interesting to note that these two equations
were used recently by Morawetz to study the quantum response of a
finite fermion system \cite{Mor00}.

Equation (\ref{frp}) combined with the equations of motion (\ref{dR})
and (\ref{dP}) gives an explicit and exact solution of the time
evolution of the Wigner function \cite{Won82}.  The equation of motion
for $\bbox{R}$ is the same as the classical equation of motion.  The
equation of motion for $\bbox{P}$ coincides with the classical
equation of motion for small $\bbox{s}$ but differs from the classical
equation when $\bbox{s}$ is large.  The latter difference
distinguishes a quantum system from a classical system.
 
We note that if (1) we expand Eq.\ (\ref{dP}) in $\bbox{s}$ and we
keep only the lowest order term, or if (2) we take the limit $\hbar
\to 0$ so that the dominant contributions in Eq.\ (\ref{frp}) come
from regions of small $\bbox{s}$, then Eqs.\ (\ref{dR}) and (\ref{dP})
becomes the classical equations of motion \cite{Won82}
\begin{eqnarray}
\label{class}
{\partial \bbox{R}\over  \partial t} = {\bbox p \over m}
\end{eqnarray}
and
\begin{eqnarray}
\label{class1}
{\partial \bbox{P} \over \partial t}
=-
\nabla_R { V(R,t)}.
\end{eqnarray}
The function $\bbox{P}(\bbox{r}_0 \bbox{p}_0 \bbox{s},t)$ is then
independent of $\bbox{s}$, and Eq.\ (\ref{frp}) gives
\begin{eqnarray}
\label{clasd}
 f (\bbox{r} \bbox{p},t)
= { \int}  d{\bbox{r}_0} d {\bbox{p}_0}
\delta [\bbox{r}-\bbox{R}(\bbox{r}_0 \bbox{p}_0,t)]
\delta[\bbox{p}-\bbox{P}(\bbox{r}_0 \bbox{p}_0,t)]
f (\bbox{r}_0 \bbox{p}_0,t_0),
\end{eqnarray}   
which is the same as solution of the evolution of the classical
distribution function. 

\section{Implementation of the Explicit Solution}

The explicit solution of Eqs. (\ref{frp}), (\ref{dR}) and (\ref{dP})
given in the previous section allows one to propagate the Wigner
function from one time to another.  One considers a small time
increment $\delta t =t-t_0$, and solves Eqs.\ (\ref{dR}) and
(\ref{dP}).  One obtains
\begin{eqnarray} 
\label{rr}
\bbox{r}=\bbox {R}(\bbox{r}_0 \bbox{p}_0 \bbox{s},t)=\bbox{r}_0+{\bbox{p}
\over m} \delta t. 
\end{eqnarray}
and 
\begin{eqnarray}
\label{pp}
\bbox{P}=\bbox{p}_0
+\bbox{e}_s 
{   V \left ( \bbox {r}- {\bbox{s} \over 2},t \right )
  - V \left ( \bbox {r}+ {\bbox{s} \over 2},t \right ) \over s} ~\delta t,
\end{eqnarray}
where $\bbox{e}_s$ is the unit vector in the $\bbox{s}$ direction.
After $\bbox{r}(\bbox{R})$ and $\bbox{P}$ for different $\bbox{s}$ have been
obtained, we can substitute them in Eq.\ (\ref{frp}) and carry out the
integration to give the new Wigner function at the next
time step.  Eqs.\ (\ref{rr}) and (\ref{pp}) lead to
\begin{eqnarray}
\label{frpt}
\fl
 f (\bbox{r} \bbox{p},t)
= { \int}  d{\bbox{r}_0}  d {\bbox{p}_0} 
\int { d \bbox{s}   \over (2 \pi\hbar)^3} 
\exp \left [  i\left \{ {\bbox{s}\cdot (\bbox{p}-\bbox{p}_0) \over \hbar}
-
   \left (   V  ( \bbox {r}- {\bbox{s} \over 2},t  )
  - V  ( \bbox {r}+ {\bbox{s} \over 2},t  )\right ) 
 {\delta t \over \hbar} \right \} \right ]
\nonumber\\
\times \delta (\bbox{r}-\bbox{r}_0-{\bbox{p} \over m}\delta t)   
f (\bbox{r}_0, \bbox{p}_0,t_0).
\end{eqnarray}                                                                 
We shall first consider the one-dimensional case for which the factor
$(2\pi \hbar)^3$ becomes $2\pi \hbar$.  The integration over
${x}_0(\bbox{r}_0)$ can be easily carried out.  the above equation for the
one-dimensional case becomes
\begin{eqnarray}
\fl
\label{fcos}
 f (x p,t)
= { \int}    d {{p}_0} 
~2\int_0^\infty { d {s}   \over 2 \pi \hbar} 
\cos \left [  {{s}\cdot ({p}-{p}_0) \over \hbar} -
\left ( V({x}-{ {s} \over 2},t)-V({x}+{ {s}
\over 2},t)  \right ) 
{\delta t \over \hbar} \right ]
\nonumber \\
\times f ({x}-{{p} \over m}\delta t, {p}_0,t_0).
\end{eqnarray}                                                                 
This explicit solution of the time evolution of the Wigner function can
be written as
\begin{eqnarray}
\label{ex1}
\fl
 f ({x} {p},t)
=  \int d {{p}_0}
\biggl \{ 
F_c({x}, {p}-{p}_0)   
+F_s({x}, {p}-{p}_0)   \biggr \}
f ({x}-{{p}\over m}\delta t, {p}_0,t_0),
\end{eqnarray}                                                                 
where 
$F_c({x}, {p}-{p}_0)$ and   
$F_s({x}, {p}-{p}_0)$
are cosine and sine
transforms given by
\begin{eqnarray}
\label{ex2}
\fl
F_c({x}, {p}-{p}_0)
=
2\int_0^\infty  { d{s}  \over 2 \pi \hbar} 
\cos\left [ {{s}\cdot ({p}-{p}_0) \over \hbar} \right ]
\cos \biggl [\biggl ( V({x}-
{ {s} \over 2},t)-V({x}+{  {s} \over 2},t)  
\biggr ) {\delta t \over \hbar}\biggr ]
\end{eqnarray}                                                                 
and
\begin{eqnarray}
\label{ex3}
\fl
F_s({x}, {p}-{p}_0) 
=
2\int_0^\infty  { d{s}  \over 2 \pi \hbar} 
\sin \left [ {{s}\cdot ({p}-{p}_0) \over \hbar } \right ]
\sin \biggl [\biggl ( V({x}-
{{s} \over 2},t)-V({x}+{ {s} \over 2},t)  
\biggr ) {\delta t \over \hbar}\biggr ].
\end{eqnarray}                                                                 
The above set of equations (\ref{ex1})-(\ref{ex3}) provide a simple
implementation of the explicit solution (\ref{frp}), (\ref{dP}), and
(\ref{dR}) of the time evolution of a one-dimensional Wigner function.

If one keeps only terms up to first order in $\delta t$ for small
values of $\delta t$, the Wigner function evolution equation becomes
\begin{eqnarray}
\label{ex4}
\fl
f ({x} {p},t)
=  
f ({x}-{{p}\over m}\delta t, {p},t_0)
+ { \delta t \over \hbar}
\int d {{p}_0}
V_s({x}, {p}-{p}_0) 
f ({x}-{{p}\over m}\delta t, {p}_0,t_0),
\end{eqnarray}                                                                 
where
\begin{eqnarray}
\label{ex5}
\fl
V_s({x}, {p}-{p}_0)
=
2\int_0^\infty { d{s}  \over (2 \pi \hbar)^3} 
\sin \left [ {{s}\cdot ({p}-{p}_0) \over \hbar } \right ]
\biggl \{V({x}-
{ {s} \over 2},t)-V({x}+{  {s} \over 2},t)  
\biggr \}.
\end{eqnarray}

We shall now discuss the three-dimensional case for which the
integration over $\bbox{s}$ in Eq.\ (\ref{frpt}) is rather
complicated.  It can be simplified if one makes the following
approximation
\begin{eqnarray}
   V  ( \bbox {r}- {\bbox{s} \over 2},t  )
  - V  ( \bbox {r}+ {\bbox{s} \over 2},t  ) 
\approx
\sum_{j=1}^3 
\left (   V  ( \bbox {r}- {s_j \bbox{e}_j \over 2},t  )
  - V  ( \bbox {r}+ {s_j \bbox{e}_j \over 2},t  )\right ),
\end{eqnarray} 
where the righthand side contains all terms of the lefthand side
except the cross terms involving products of different $s_j$ and
$s_{j'}$$(j' \ne j)$ with high-order derivatives of $V(\bbox{r})$,
when one expand the above functions in powers of $s_j$.  If one uses
this approximation by neglecting these cross terms, Eq.\ (\ref{frpt})
for the three-dimensional case becomes
\begin{eqnarray}
\fl
\label{fcos3}
 f (\bbox{r} \bbox{p},t)
= { \int}    d {\bbox{p}_0} 
\prod_{j=1}^3 ~2\int_0^\infty { d {s}_j   \over 2 \pi \hbar} 
\cos \left [  {{s}_j({p}_j-{p}_{0j}) \over \hbar} -
\left ( V(\bbox{r}-{s_j\bbox {e}_j \over 2},t)-V(\bbox{r}+{ s_j\bbox{e}_j
\over 2},t)  \right ) 
{\delta t \over \hbar} \right ]
\nonumber \\
\times f (\bbox{r}-{\bbox{p} \over m}\delta t, \bbox{p}_0,t_0).
\end{eqnarray}                                                                 
The explicit solution of the time evolution of the Wigner
function can be written as
\begin{eqnarray}
\label{ex13}
\fl
 f (\bbox{r} \bbox{p},t)
=  \int d {\bbox{p}_0}
\prod_{j=1}^3 \biggl \{ 
F_{cj}(\bbox{r}, {p_j}-{p}_{0j})   
+F_{sj}(\bbox{r}, {p_j}-{p}_{0j})   \biggr \}
f (\bbox{r}-{\bbox{p}\over m}\delta t, \bbox{p}_0,t_0),
\end{eqnarray}                                                                 
where 
$F_{cj}(\bbox{r}, {p_j}-{p}_{0j})$ and   
$F_{sj}(\bbox{r}, {p_j}-{p}_{0j})$
are cosine and sine
transforms given by
\begin{eqnarray}
\label{ex23}
\fl
F_{cj}(\bbox{r}, {p_j}-{p}_{0j})
=
2\int_0^\infty  { d{s}_j  \over 2 \pi \hbar} 
\cos\left [ {{s}_j ({p_j}-{p}_{0j}) \over \hbar} \right ]
\cos \biggl [\biggl ( V(\bbox{r}-
{ s_j \bbox{e}_j \over 2},t)-V(\bbox{r}+{ s_j \bbox{e}_j \over 2},t)  
\biggr ) {\delta t \over \hbar}\biggr ]
\nonumber\\
\end{eqnarray}                                                                 
and
\begin{eqnarray}
\label{ex33}
\fl
F_{sj}(\bbox{r}, {p_j}-{p}_{0j}) 
=
2\int_0^\infty  { d{s}_j  \over 2 \pi \hbar} 
\sin \left [ {{s_j} ({p_j}-{p}_{0j}) \over \hbar } \right ]
\sin \biggl [\biggl ( V(\bbox{r}-
{ s_j \bbox{e}_j \over 2},t)-V(\bbox{r}+{ s_j \bbox{e}_j \over 2},t)  
\biggr ) {\delta t \over \hbar}\biggr ].
\nonumber\\
\end{eqnarray}                                                                 
The above set of equations (\ref{ex13})-(\ref{ex33}) provide a simple
and approximate implementation of the explicit solution (\ref{frp}),
(\ref{dP}), and (\ref{dR}) of the time evolution of a
three-dimensional Wigner function.

\section{An Example of the Evolution of the Wigner Function}

It is of interest to consider an explicit example to test whether the
solutions in Eqs.\ (\ref{ex1})-(\ref{ex3}) or (\ref{ex4})-(\ref{ex5})
leads to the correct results.  For this purpose we study the dynamics
of the Wigner function for a case that can be easily evaluated and
compared with the Wigner function obtained by using the explicit
solution.

We examine a particle in a one-dimensional attractive Gaussian
potential well.  We express physical quantities in dimensionless
units, with a Schr\"odinger equation given by
\begin{eqnarray}
H\Psi(x) =\left \{- {1\over 2} {d^2 \over dx^2} 
-e^{-{ x^2 /2 \sigma^2}}\right \} \Psi(x).
\end{eqnarray}
The wave function $\Psi(x)$ can be expanded
in terms of a set of non-orthogonal Gaussian wave functions with
different widths,
\begin{eqnarray}
\label{psi}
\Psi (x) = \sum_{n=1}^{n_{\rm max}} a_n \psi_n (x),
\end{eqnarray}
where the normalized basis function $\langle x | n \rangle=\psi_n(x)$
is taken to be
\begin{eqnarray}
\psi_n(x) ={ e^{-x^2/2\beta_n^2} \over (\sqrt{\pi}\beta_n)^{1/2}},
\end{eqnarray}
and $\beta_n^2=n \beta_0^2$.  The Hamiltonian matrix in this basis set
can be easily constructed.  Specifically, one calculates the matrix
element of the overlap matrix $B$ for this set of basis states
\begin{eqnarray}
B_{nm}=\langle n | m \rangle 
= \sqrt{2 \beta_n\beta_m \over \beta_n^2 + \beta_m^2},
\end{eqnarray}
the matrix element of the kinetic energy matrix $T$
\begin{eqnarray}
T_{nm}=\langle n |\left \{ - {1\over 2} {d^2 \over dx^2}\right \}| m
\rangle 
= {1 \over
2} \sqrt{2 \beta_n\beta_m \over \beta_n^2 
+ \beta_m^2}{1 \over \beta_n^2 + \beta_m^2} ~,
\end{eqnarray}
and the matrix element for potential energy matrix $V$
\begin{eqnarray}
V_{nm}=\langle n |(-e^{-{ x^2 /2 \sigma^2}})|m\rangle
=-\sqrt{ {2 \beta_n\beta_m \over \beta_n^2 + \beta_m^2}
{\sigma^2 \over \beta_{nm}^2 +\sigma^2}}~,
\end{eqnarray}
where
\begin{eqnarray}
\beta_{nm}^2={\beta_n^2 \beta_m^2 \over \beta_n^2+\beta_m^2}.
\end{eqnarray}
The eigenvalue equation becomes
\begin{eqnarray}
(T+V)a=EBa,
\end{eqnarray}
where $a$ is the column matrix of the coefficients $\{a_n\}$ of Eq.\
(\ref{psi}).  This eigenvalue equation can be diagonalized to yield
the eigenenergies $E_\lambda$ and eigenfunctions $\Psi_\lambda (x)$
represented by the set of coefficients $\{a_{\lambda n} \}$.  We
normalize $\Psi_\lambda (x)$ by $\int dx |\Psi_\lambda (x)|^2 =1$.

A simple non-stationary state $\Phi(x)$ of the system can be
constructed as a linear combination of the eigenstates $\Psi_\lambda
(x)$ with amplitudes $b_\lambda$,
\begin{eqnarray}
\Phi(x,t=0) = \sum_\lambda b_\lambda \Psi_\lambda (x).
\end{eqnarray}
The time dependence of this non-stationary  system is 
\begin{eqnarray}
\Phi(x,t) = \sum_\lambda b_\lambda e^{-iE_\lambda t} \Psi_\lambda (x).
\end{eqnarray}
The Wigner function of the system is then given by
\begin{eqnarray}
\label{time}
f(xp,t) = \sum_{nm} c_n(t) c_m^*(t) f_{nm}(xp),
\end{eqnarray}
where
\begin{eqnarray}
c_n(t) = \sum_\lambda b_\lambda e^{-iE_\lambda t} a_{\lambda n},
\end{eqnarray}
\begin{eqnarray}
f_{nm}(xp) = 2 \sqrt{2 \beta_n\beta_m \over \beta_n^2 + \beta_m^2} 
e^{-\nu_{nm}},
\end{eqnarray}
\begin{eqnarray}
\nu_{nm}={(\beta_n^2+\beta_m^2 ) 
(x^2 - \mu_{nm}^2/4) \over 2 \beta_n^2\beta_m^2} ,
\end{eqnarray}
and
\begin{eqnarray}
\mu_{nm}=4 {\beta_n^2\beta_m^2 \over \beta_n^2+\beta_m^2 }
 \left [ x \left ( {1 \over 2 \beta_n^2 }- {1\over 2\beta_m^2}\right )
 + ip \right ] ~.
\end{eqnarray}

In our numerical example, we study the dynamics of a system in a
potential with a width parameter $\sigma=3$.  The two lowest energy
even-parity eigenstates $\Psi_0(x)$ and $\Psi_1(x)$ have eigenvalues
$E_0= -0.844$ and $E_1=-0.312$ respectively.  As a test, we construct
a non-stationary wave function with equal amplitudes in these two
eigenstates at $t=0$,
\begin{eqnarray}
\label{non}
\Phi(x,t=0) = {1 \over \sqrt{2}} \{\Psi_0 (x) +\Psi_1(x)\}.
\end{eqnarray}
The time dependence of the wave function is then
\begin{eqnarray}
\Phi(x,t) = {1 \over \sqrt{2}} \{e^{-iE_0 t}\Psi_0 (x) +e^{-iE_1 t}\Psi_1(x)\}.
\end{eqnarray} 
The time dependence of the Wigner function can be evaluated using Eq.\
(\ref{time}).  The Wigner functions for different values of $p$ and
$t$ are shown in Figs.\ (1a), (1b), and (1c) on the left panel of
Fig.\ 1.  The Wigner function has both positive and negative values in
different regions of phase space.  The eigenfunctions and eigenvalues
give a correct representation of the Wigner function as it depends
only on the accuracy of the eigenvalues and eigenfunctions, and not on
the accuracy of the method of propagating forward in time.

\begin{figure}[h]
\begin{center}
\epsfig{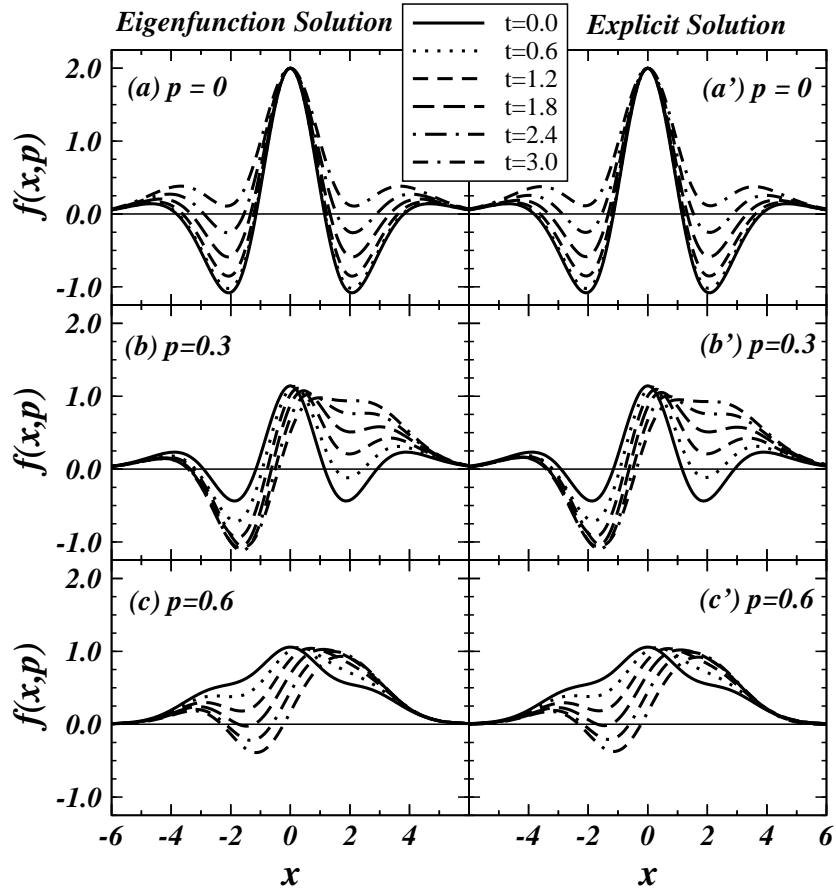}
\end{center}
\caption{Time dependence of the Wigner function obtained in two
different methods. The left panel gives the Wigner function obtained
from eigenfunctions and eigenvalues, and the right panel shows that
Wigner function obtained by the explicit solution of Eqs.\
(\ref{ex1})-(\ref{ex3}).  }
\end{figure}

The non-stationary state we have used leads to an initial Wigner
function with significant spatial and temporal variations as well as
regions of negative values.  These properties provide a stringent test
of the methods of using Eqs.\ (\ref{ex1})-(\ref{ex3}) or Eqs.\
(\ref{ex4})-(\ref{ex5}) to propagate the Wigner function.

For our tests we take the initial Wigner function for our
non-stationary state Eq.\ (\ref{non}) at time $t=0$, and evolve the
Wigner function using the explicit solution.  We carry out the time
evolution using small time increments.  We can choose either Eqs.\
(\ref{ex1})-(\ref{ex3}) or Eqs.\ (\ref{ex4})-(\ref{ex5}) to evolve the
Wigner function.  The former includes higher order effects in $\delta
t$, and can be used with a larger time increment than the second
approach.  We choose to use Eqs.\ ({\ref{ex1})-(\ref{ex3}), and the
results for the time evolution from $t=0$ to $t=3$ in Figs.\ (1a'),
(1b'), and (1c') were calculated using 30 time steps.  The cosine and
sine transforms can be calculated using the method of fast Fourier
transforms.  The numerical integration over $p_0$ in Eq.\ (\ref{ex1})
can then be carried out to yield the Wigner function at the next time
step.  This process is repeated in a stepwise manner to propagate
forward in time.  The resulting time evolution of the Wigner function
is shown in Figs.\ (1a'), (1b'), and (1c') on the right panel of Fig.\
1.

A comparison of the left and the right panels indicates that the
explicit solution of Eqs.\ (\ref{ex1})-(\ref{ex3}) gives an excellent
reproduction of the time evolution of the Wigner function obtained by
using eigenvalues and eigenfunctions.  The phases and the negative
regions of the Wigner function are correctly reproduced.  The
difference between the eigenfunction solution and the explicit
solution is small.  For example, for $p=0.6$ at $t=3$, the maximum
value of the eigenfunction solution of the Wigner function is 0.9313
at $x=1.692$, and the explicit solution gives 0.9206 at $x=1.692$. For
this momentum $p$ and $t=3$, the minimum of the eigenfunction solution
of the Wigner function is $-0.3888$ at $x=-1.1667$, and the the explicit
solution gives $-0.3735$ at $x=-1.1667$.  The positions of the maxima
and minima in the two methods are the same, and the magnitudes differ
by about 1\%.  The explicit solution of Eqs.\ (\ref{ex1})-(\ref{ex3})
thus leads to an accurate determination of the time evolution of the
Wigner function, even if it contains regions of negative or
oscillating values.

\section{Pseudoparticle Method and Approximations} 
As an approximation to obtain the explicit solution of
Eqs.\ (\ref{frp}), (\ref{dR}) and (\ref{dP}), a pseudoparticle method
was presented previously to calculate the time evolution of the Wigner
function \cite{Won82}.  It is useful to test the pseudoparticle method
with our numerical example to assess its usefulness and accuracy.

To introduce the pseudoparticle method, we rewrite Eq.\
(\ref{frpt}) in the form \cite{Won82,note}
\begin{eqnarray}
\label{dd}
\fl 
~~~~
f (\bbox{r} \bbox{p},t) = { \int} d{\bbox{r}_0} d {\bbox{p}_0}
\Delta(\bbox{p}-\bbox{p}_0+  \nabla_r V(\bbox{r},t) \delta t) \delta
(\bbox{r}-\bbox{r}_0-{\bbox{p} \over m}\delta t) f (\bbox{r}_0,
\bbox{p}_0,t_0),
\end{eqnarray}  
where
\begin{eqnarray} 
\label{Delta}
\fl
~~~~
\Delta(\bbox{\pi})
=\int { d \bbox{s}   \over (2 \pi\hbar)^3}
\exp \left [  i\left \{ {\bbox{s}\cdot \bbox{\pi} \over \hbar}
+{2 \delta t \over \hbar}   
\sum_{n=1}^{\infty} {1 \over (2n+1)!} 
\left ({\bbox{s}\cdot \nabla_r^V \over 2} \right )^{2n+1}
V(\bbox{r},t)  \right \}  \right ],
\end{eqnarray}  
and $\bbox{\pi}=\bbox{p}-\bbox{p}_0+ \nabla_r V(\bbox{r},t) \delta t $.  One
notes that the second term in the exponential function is a sum
involving third and higher powers of $\bbox{s}$.  If one neglects 
these higher-order terms in the exponential function, then one obtains
the approximation
\begin{eqnarray}
\Delta (\bbox{p}-\bbox{p}_0+ \nabla_r V(\bbox{r},t)  \delta t )
\approx
\delta (\bbox{p}-\bbox{p}_0+ \nabla_r V(\bbox{r},t)  \delta t )
\end{eqnarray}
and the time evolution of the Wigner function becomes \cite{Won82}
\begin{eqnarray}
\label{psu0}
\fl
 f (\bbox{r} \bbox{p},t)\approx
 f_{LO} (\bbox{r} \bbox{p},t)
= { \int}  d{\bbox{r}_0}  d {\bbox{p}_0}
\delta (\bbox{p}-\bbox{p}_0+ \nabla_r V(\bbox{r},t) \delta t)
\delta (\bbox{r}-\bbox{r}_0-{\bbox{p} \over m}\delta t)
f (\bbox{r}_0, \bbox{p}_0,t_0)
\end{eqnarray} 
which leads to 
\begin{eqnarray}
\label{psu}
 f (\bbox{r} \bbox{p},t)\approx
f_{LO} (\bbox{r} \bbox{p},t) = 
f (\bbox{r}-\bbox{p}\delta t /m,~ \bbox{p}+\nabla_r V(\bbox{r},t) \delta t,~~t_0).
\end{eqnarray} 
The above equation provides the basis for the lowest-order (LO)
pseudoparticle approximation of the time evolution of the Wigner
function \cite{Won82}.  One divides the phase space into cells
(pseudoparticles) and follows the trajectories of these
pseudoparticles using classical equations of motion, Eqs.\
(\ref{class}) and (\ref{class1}).  Eq.\ (\ref{psu}) specifies that the
Wigner function at the new phase space point $(\bbox{r}\bbox{p})$ at
$t=t_0+\delta t$ is approximately the same as the initial Wigner
function at the initial classical trajectory phase space point
$(\bbox{r}-\bbox{p} \delta t /m, ~\bbox{p}+\nabla_r V(\bbox{r},t)
\delta t)$ at $t_0$.  The subscript $LO$ in $f_{LO} (\bbox{r}
\bbox{p},t)$ is to indicate that it is the solution of the
lowest-order pseudoparticle approximation obtained by following
classical trajectories.  This method was used in \cite{Won82} where it
was found that the lowest-order pseudoparticle approximation
reproduces quite well the general features of the time-dependent
Hartree-Fock approximation.  The use of classical trajectories to
obtain an approximate time evolution of the Wigner function was also
emphasized by Zachos and Curtright \cite{Cur99}.

\begin{figure}[h]
\begin{center}
\epsfig{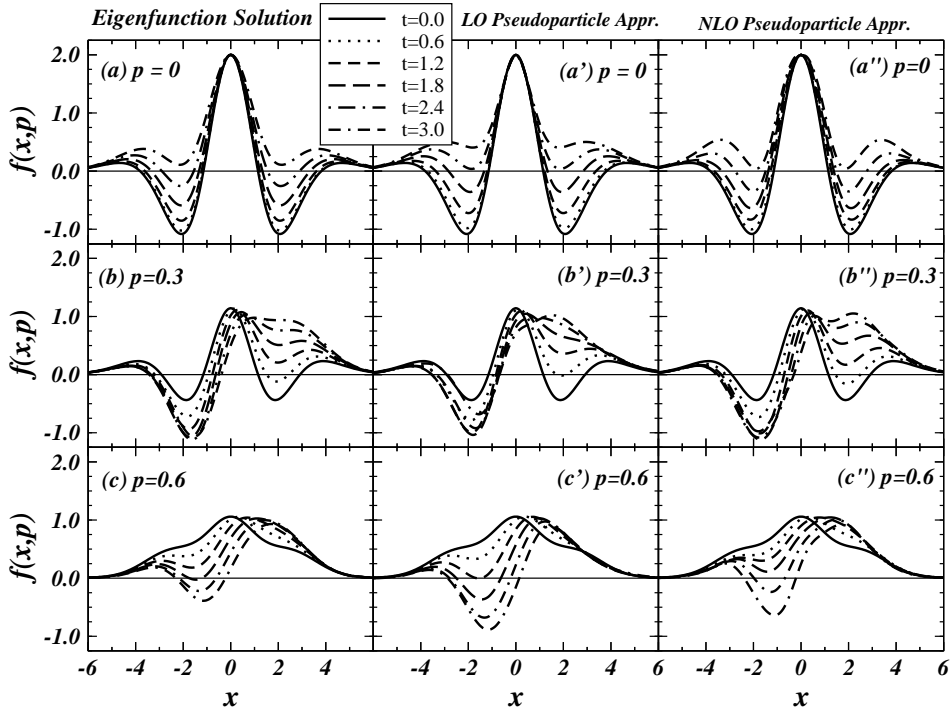}
\end{center}
\caption{Time dependence of the Wigner function obtained in different
methods. The left panel gives the correct Wigner function obtained
from eigenfunctions and eigenvalues, the middle panel gives the
results from the lowest-order (LO) pseudoparticle approximation, and
the right panel shows the Wigner function obtained by the
next-to-lowest (NLO) order pseudoparticle approximation with $\hbar^2$
corrections.  }
\end{figure}                                                             

How good is the lowest-order pseudoparticle approximation?  The
accuracy of the pseudoparticle approximation depends on the potential
$V(\bbox{r},t)$.  If the potential is a constant, a linear, or a
harmonic oscillator potential, then even the LO pseudoparticle method
gives the exact solution, as is clearly indicated in Eq.\
(\ref{Delta}).  The LO pseudoparticle solution deviates from the correct
solution if the third and higher order spatial derivatives of the
potential do not vanish.

For the non-stationary state in a Gaussian potential in our test, we
show the results of the LO pseudoparticle approximation in Figs.\
(2a'), (2b'), and (2c'), to be compared with the correct results from
the eigenfunction method in Figs. (2a), (2b), and (2c).  In our
calculations, we specify the Wigner function at phase space
coordinates of a fixed lattice.  The Wigner function at the phase
space points on the righthand side of Eq.\ (\ref{psu}) is obtained by
interpolation using the Wigner function $f(\bbox{r}_0,\bbox{p}_0,t_0)$
at the earlier time $t_0$.  As one observes, the general features of
the oscillations are approximately reproduced.  However, there are
deviations from the correct results as time increases.  For example,
the Wigner function values for $p=0$, $t=3$, and $|x| \sim 2$ obtained
in the LO pseudoparticle approximation in Fig.\ (2a') are much larger
than the corresponding correct values in Fig.\ (2a).  The values of
the Wigner function for $p=0.6$, $t=3$, and $x \sim -1$ in the LO
pseudoparticle approximation in Fig.\ (2c') are much smaller than the
corresponding correct values in Fig.\ (2c).

We can correct for the deviations of the LO pseudoparticle
approximation.  For a small time increment $\delta t$, we can keep
terms up to the first order in $\delta t$ in Eq.\ (\ref{Delta}).  The
function $\Delta(\bbox{\pi})$ becomes
\begin{eqnarray}
\fl
~~~~
\Delta(\bbox{\pi}) 
=\int { d \bbox{s}   \over 2 \pi\hbar}
e^{   i \bbox{s}\cdot \bbox{\pi} / \hbar} 
\left [ 1+{2i \delta t \over \hbar} \sum_{n=1}^{\infty} {1 \over (2n+1)!}
\left ({\bbox{s}\cdot \nabla_r^V \over 2} \right )^{2n+1}
V(\bbox{r},t)\right ].
\end{eqnarray} 
The integration over $\bbox{s}$ can be easily carried out and we
obtain
\begin{eqnarray}
\label{delta}
\fl
~~~~~~~~
\Delta(\bbox{\pi}) 
=\left [ 1
+\delta t
\sum_{n=1}^{\infty} {1 \over (2n+1)!}
\left ({\hbar\over 2i}\right )^{2n} ( \nabla_r^V \cdot \nabla_p^\delta)^{2n+1}
V(\bbox{r},t) \right ] \delta(\bbox{\pi}),
\end{eqnarray} 
where $\nabla_r^V$ acts on $V(\bbox{r},t)$ and $\nabla_p^\delta$ acts
on $\delta(\bbox{\pi})$.  
The corrected pseudoparticle solution of the
time evolution of the Wigner function is then
\begin{eqnarray}
\label{psucor}
\fl 
~~~~~~~f (\bbox{r} \bbox{p},t)
=\left [ 1 +\delta t
\sum_{n=1}^{\infty} {1 \over (2n+1)!}
\left ({\hbar\over 2i} \right )^{2n} 
( \nabla_r^V \cdot \nabla_p^f)^{2n+1}
V(\bbox{r},t) \right ] f_{LO}(\bbox{r}, \bbox{p},t),
\end{eqnarray} 
where $\nabla_p^f$ acts on $f_{LO}(\bbox{r}, \bbox{p},t)$.  It should be
emphasized that no odd powers of $\hbar$ and even powers of $(
\nabla_r^V \cdot \nabla_p^f)$ are present in the above expansion.  We
can improve upon the pseudoparticle approximation by including
additional contributions in powers of $\hbar^2$ involving higher-order
derivatives of $V(\bbox{r},t)$ and $f_{LO}(\bbox{r}, \bbox{p},t)$.

We show in Figs.\ (2a''), (2b''), and (2c'') the results in the
next-to-lowest order approximation obtained by including the
correction term of order $\hbar^2$, which involves the third
derivatives of the potential and the Wigner function.  In this
calculation, we propagate the Wigner function from $t_0$ to
$t=t_0+\delta t$ for a small time increment $\delta t$ following
classical trajectories as in Eq.\ (\ref{psu0}).  The Wigner function
is then corrected at time $t$ using Eq.\ (\ref{psucor}).  The
corrected Wigner function at $t$ is used in Eq.\ (\ref{psu0}) to
propagate to time $t+\delta t$ by following classical trajectories.
It is then corrected at time $t+\delta t$ using Eq.\ (\ref{psucor}).
These steps are repeated to obtain the time evolution of the Wigner
function in this pseudoparticle approximation.  To avoid numerical
instability in the time evolution, we use the Wigner function
$f_{LO}(\bbox{r} \bbox{p},t)$ obtained in the lowest-order
approximation to calculate the third momentum derivative, when we
evaluate the $\hbar^2$ correction term in Eq.\ (\ref{psucor}).

Results in Fig.\ 2 indicate that the inclusion of the $\hbar^2$
correction term in the next-to-lowest (NLO) order improves the
pseudoparticle approximation.  In particular, the NLO values of the
Wigner function at $p=0$, $t=3$, and $|x| \sim 2$ obtained with
$\hbar^2$ corrections Fig.\ (2a'') are now closer to the values
obtained with eigenfunctions.  The NLO values of the Wigner function
at $p=0.6$, $t=3$, and $x \sim -1$ in Fig.\ (2c') are also closer to
the corresponding values obtained with eigenfunctions.  Although small
deviations remain in the NLO solution, the general features of the
Wigner function are reasonably well reproduced.

We give here an alternative method to evaluate the correction terms in
the pseudoparticle method.  As one notes in Eq.\ (\ref{delta}), the
correction terms involve high-order derivatives of the
$\delta$-function.  They can be evaluated approximately by expressing
the $\delta$-function in terms of known functions.  Making use of the
othonormality of the harmonic oscillator wave functions
$\{\phi_m(x)\}$, we can represent a one-dimensional delta function
$\delta(x)$ by
\begin{eqnarray}
{ \delta}(x)
=\sum_{m=0}^\infty \phi_m^*(x) \phi_m(0)=
{\alpha \over \sqrt{\pi}}
e^{-{\alpha^2 x^2 \over 2}} \sum_{m=0}^\infty H_{2m}(\alpha x){(-1)^m \over
 2^{2m} m!},
\end{eqnarray}
where $H_{2m}(\alpha x)$ is the Hermite polynomial of order $2m$, and
$\alpha$ is a width parameter.  In practice, the summation in the
above equation is carried out up to an order $M$, and the delta
function $\delta(x)$ can be approximated by the distribution
\begin{eqnarray}
\label{DD}
\delta(x) \approx D(x)
=A_M {\alpha \over \sqrt{\pi}}
e^{-{\alpha^2 x^2 \over 2}} \sum_{m=0}^M H_{2m}(\alpha x){(-1)^m \over
 2^{2m} m!},
\end{eqnarray}
where $A_M$ is
\begin{eqnarray}
A_M=\left [ \sqrt{2} \sum_{m=0}^M (-1)^m {(2m)!\over 2^{2m}m!m!}
  \right ]^{-1},
\end{eqnarray}
and $A_M$ is so chosen that $D(x)$ is normalized to $\int D(x)dx =1$.
A three-dimensional generalization of this $D$-function,
$D(\bbox{x})$, can be obtained as the product of three one-dimensional
$D$-functions of their respective component coordinates,
$D(x_x)D(x_y)D(x_z)$.  Using a distribution function of this type, we
can write the explicit solution of the Wigner function as
\begin{eqnarray}
\label{psucor1}
\fl 
~~f (\bbox{r} \bbox{p},t)
=
 f_{LO}(\bbox{r}, \bbox{p},t)
+ \delta t 
{ \int} d{\bbox{r}_0}  d {\bbox{p}_0}
\delta (\bbox{r}-\bbox{r}_0-{\bbox{p} \over m}\delta t)
f (\bbox{r}_0,\bbox{p}_0,t_0)
\nonumber\\
%
\!\!\!\!\!\!\!\!\!\!\!\!\!
\times
\sum_{n=1}^{\infty} {1 \over (2n+1)!}
\left ({\hbar\over 2i} \right )^{2n} 
\left [ ( \nabla_r^V \cdot \nabla_p^D)^{2n+1}
V(\bbox{r},t)  D(\bbox{p}-\bbox{p}_0+\delta t \nabla_r V(\bbox{r},t))
\right ],
\end{eqnarray} 
where $\nabla_p^D$ applies to $D(\bbox{p}-\bbox{p}_0+\delta t\nabla_r
V(\bbox{r},t))$.  As $D$ is a known analytical function of $\bbox{p}$,
$(\nabla_p^D)^{2n+1} D(\bbox{p}-\bbox{p}_0+\delta t \nabla_r
V(\bbox{r},t))$ can be obtained analytically.  The higher-order
derivatives of the Wigner function in Eq.\ (\ref{psucor}) can be
converted into an integration over $\bbox{p}_0$.

The pseudoparticle method utilizes classical trajectories and is easy
to use.  It is even exact for a constant, linear, or harmonic
oscillator potential.  The development of systematic higher-order
corrections to the pseudoparticle method for a general potential in
Eq.\ (\ref{psucor}) allows one to improve on the approximate solution
to make the pseudoparticle method a useful tool for future
applications.

In applying the pseudoparticle method, one can use the Eulerian
picture with a fixed lattice or alternatively the Lagrangian picture
with a set of pseudoparticle phase space coordinates, as in molecular
dynamics.  The distribution $D$ of Eq.\ (\ref{DD}) can be used to go
from one picture to the other.  For example, if one starts with a
Wigner function in the Eulerian picture, one divides the phase space
into pseudoparticle cells with volume element $\Delta \bbox{r}_i
\Delta \bbox{p}_i$ and specifies the phase space coordinates
$(\bbox{r}_i \bbox{p}_i)$ and its Wigner function $f(\bbox{r}_i
\bbox{p}_i)$.  The set of variables $(\bbox{r}_i \bbox{p}_i)$ and
$f_L(\bbox{r}_i \bbox{p}_i)$$(=f(\bbox{r}_i \bbox{p}_i))$ can be used
in the Lagrangian picture to evolve the phase space dynamics.
Conversely, if one is given the set of variables $(\bbox{r}_i
\bbox{p}_i)$ and $f_L(\bbox{r}_i \bbox{p}_i)$ in the Lagrangian
picture, then the Wigner function in the Eulerian picture
$f_E(\bbox{r}\bbox{p})$ at coordinates $(\bbox{r}\bbox{p})$ is given
by
\begin{eqnarray}
\label{Eul}
f_E(\bbox{r}\bbox{p})=\sum_{i} D_r(\bbox{r}-\bbox{r}_i)
 D_p(\bbox{p}-\bbox{p}_i)
 f_L(\bbox{r}_i \bbox{p}_i) \Delta \bbox{r}_i  \Delta \bbox{p}_i, 
\end{eqnarray}
where the width parameter $\alpha$ in $D_r$ (and $D_p$) depends on the
magnitude of $\Delta \bbox{r}_i$ (and $\Delta \bbox{p}_i$).  These
transcriptions between the two pictures facilitate the application of
the pseudoparticle method.  

\section{Conclusions and Discussions}

We have reviewed and applied here the explicit solution of the time
evolution of the Wigner function obtained previously in 1982
\cite{Won82}.  The basic idea is to represent the Wigner function in
terms of auxiliary phase space coordinates, which obey simple
equations of motion.  These equations are similar to the classical
equations of motion.  They can be solved easily.  The solutions of
these equations of motion can then be used to evaluate the time
evolution of the Wigner function.  We have demonstrated the usefulness
of the explicit solution using a numerical example.  We find that the
explicit solution leads to the correct time evolution of the Wigner
function, even for a Wigner function with strong spatial and temporal
variations and regions of negative values.

We have also reviewed and tested the pseudoparticle method for the
evaluation of the time evolution of the Wigner function.  For our
example of a non-stationary state in a Gaussian potential, the
lowest-order pseudoparticle approximation gives the correct features
of the time evolution, but there are deviations from the correct
results.  We have developed a systematic way to improve the
pseudoparticle method involving correction terms in powers of
$\hbar^2$ containing high-order derivatives of the potential and the
Wigner function.

The simplicity of the different methods discussed here will facilitate
their application to quantum dynamics in phase space.  They can be
used to study quantum particle dynamics in a time-dependent,
multi-dimensional potential.  They can also be applied to study the
one-body Wigner function of a many-particle system in a time-dependent
mean-field potential, as in \cite{Won82}.  With the addition of a
collision term, they can be used to describe the dynamics of a quantum
Boltzmann equation.  The explicit solution of
Eqs. (\ref{ex1})-(\ref{ex3}) or (\ref{ex4})-(\ref{ex5}) is probably
best handled in a lattice of phase space points, as the sine and
cosine transforms are simplest in such a lattice.  On the other hand,
the pseudoparticle method of Eqs.\ (\ref{psu0}) and
(\ref{psucor}) can be investigated both in terms of pseudoparticle
phase space coordinates in the Lagrangian picture or alternatively in
a fixed lattice of phase space points in the Eulerian picture.  The
use of pseudoparticle coordinates may provide substantial saving of
computer storage capacity when a significant fraction of the phase
space is empty.  Future research using these methods will allow us to
explore further the richness of quantum dynamics in phase space, which
Professor Wigner first pioneered for us.

\ack The author would like to thank Drs. E. Pollak and P. G. Reinhard
for helpful discussions and communications.  This research was
supported by the Division of Nuclear Physics, Department of Energy,
under Contract No. DE-AC05-00OR22725 managed by UT-Battelle, LLC.


\section*{References}

\begin{thebibliography}{99}

\bibitem{Wig32} E. P. Wigner, Phys. Rev. {\bf 40}, 749 (1932).

\bibitem{Wig03} 

?. Janszky, Y. S. Kim, and M. A. Man'ko, {\it Wigner function and
phase-space approach in quantum mechanics}, to be published in the
Journal of Optics B: Quantum and Classical Optics, June 2003.

\bibitem{Won82} C. Y. Wong, Phys. Rev. {\bf C25}, 1460 (1982).

\bibitem{Smi93} 
D. T. Smithey $et~al.$ Phys. Rev. Lett. {\bf 70}, 1244 (1993).

\bibitem{Lei96}
D. Leibfried $et~al.$, Phys. Phys. Rev. Lett. {\bf 77}, 4281
(1996).

\bibitem{Bri97}
G. Breitenbach, S. Schiller, and J. Mlynek, Nature {\bf 387}, 471
(1997).

\bibitem{Kur97}
Ch. Kurtsiefer, T. Pfau, and J. Mlynek, Nature {\bf 386}, 150
(1997).

\bibitem{Lvo01} 
A. I. Lvovsky $et~al.$ Phys. Rev. Lett. {\bf 87}, 050402 (2001).

\bibitem{Lou02} P. Lougovski, E. Solano, Z. M. Zhang, H. Walther,
H. Mack, and W. P. Schleich, quant-ph/0206083.

\bibitem{Joh87}
S. John and E. A. Remler,
Ann.of Phys. 180 (1987) 152.

\bibitem{Pru78}
E. Prugove\v cki, Ann. Phys.(N.Y.) {\bf 10}, 102 (1978).

\bibitem{Tak89}
K. Takahashi,
Prog. Theor. Phys. Supplement, {\bf 98}, 109 (1989).


\bibitem{Man96} S. Mancini, V. I. Man'ko, and P. Tombesi,
Phys. Lett. {\bf A213}, 1 (1996); for a review, see 
V. I. Man'ko,  quant-ph/9902079.

\bibitem{Cur99} T. Curtright and C. K. Zachos, J. Phys. {\bf A32}, 771
(1999); C. K. Zachos and T. Curtright, Prog. Theor. Phys. Suppl. {\bf
135}, 244 (1999).

\bibitem{Lev94} M. Levanda and V. Fleurov, J. Phys. : Condensed Matter 
{\bf 6}, 7889 (1994);  M. Levanda and V. Fleurov, Ann. Phys. {\bf 292}, 199
(2001);  M. Levanda and V. Fleurov, cond-mat/0111436.

\bibitem{Fil96} V. S. Filinov, J. Mol. Phys. {\bf 88}, 1517, 1529
(1996); V. Filinov, Yu. Lozovik, A. Filinov, I. Zacharov, and
A. Oparin, Physica Scripta {\bf 58}, 304 (1998); V. S. Filinov,
P. Thomas, I. Varga, T. Meier, M. Bonitz, V. Fortov, and S. Koch,
Phys. Rev. {\bf B65}, 165124 (2002), cond-mat/0203585.

\bibitem{Ank02} J. Ankerhold, M. Saltzer, and E. Pollak,
J. Chem. Phys. {\bf 116 (14)}, 5925 (2002); J. L. Liao and  E. Pollak,
J. Chem.  Phys. {\bf 116 (7)}, 2718 (2002); E. Pollak and J. S. Shao,
J. Chem.  Phys. {\bf 116 (4)}, 1748 (2002).

\bibitem{Shi02}
L. Shifren and D. K. Ferry, Physica {\bf B314}, 72 (2002).

\bibitem{Mor00}
K. Morawetz, Phy. Rev. {\bf E61}, 2555 (2000).

\bibitem{Fed01}
A. N. Fedorova and M. G. Zeitlin,
in Proceedings of the Capri ICFA Workshop, October, 2000,
physics/0101006.

\bibitem{Elz89}
H.-Th. Elze and U. Heinz, Phy. Rep. {\bf 183}, 81 (1989); H.-Th. Elze
Nucl. Phys. {\bf B436}, 213 (1995); H.-Th. Elze and J. Rafelski, and
L. Turko, Phys. Lett. {\bf B506}, 123 (2001).

\bibitem{Bla99}
J. P. Blazoit and E. Iancu, Nucl. Phys. {\bf B557}, 183 (1999).

\bibitem{Rei95}
P.-G. Reinhard and E. Suraud, Ann. Phys. (N.Y.) {\bf 216}, 98 (1995).

\bibitem{Sch96}
J. Schnack and H. Feldmeier, Nucl. Phys. {\bf A601}, 181 (1996).


\bibitem{note} Eqs.\ (2.12) and (2.13) in Ref.\ \cite{Won82}
correspond respectively to Eqs.\ (\ref{dd}) and (\ref{Delta}) in this
paper.  There were inadvertent typographical errors in the former
equations. In Eq.\ (2.12) of \cite{Won82}, $\nabla \cal V$ should read
$\nabla {\cal V} \delta t$ and $f(\bbox{r}_0,\bbox{p}_0,t)$ should
read $f(\bbox{r}_0,\bbox{p}_0,t_0)$, and in Eq. (2.13), the factor
$1/6$ should read $(1/6) \times (\delta t/4)$.

\end{thebibliography}
\end{document}